\documentclass[11pt]{article} 
\usepackage{t1enc}
\usepackage{amsmath, amssymb}
\usepackage{amsfonts}
\def\be{\begin{equation}}
\def\ee{\end{equation}}

\def\be{\begin{equation}}
\def\ee{\end{equation}}

\begin{document}
\rightline{RI-12-03}
\rightline{FNT/T 2003/15}
\rightline{hep-th/0312186} \rm \vskip 0.3in

\begin{center}
{\Large \textbf{\ Holography in asymptotically flat space-times
and the BMS group}}

\vspace{1 cm}

{\bf Giovanni Arcioni}$^{a,}$\footnote{ E-mail :
arcionig@phys.huji.ac.il} and {\bf Claudio
Dappiaggi}$^{b,}$\footnote{ E-mail :
claudio.dappiaggi@pv.infn.it},

\vspace{1cm}

$^{a}$ {\it The
Racah Institute of Physics, The Hebrew University, \\[0pt] Jerusalem 91904, Israel.}

\vspace{0.1cm}

$^{b}${\it ~Dipartimento di Fisica Nucleare e Teorica,

Universit\`{a} degli Studi di Pavia, INFN, Sezione di Pavia, \\[0pt]
via A. Bassi 6, I-27100 Pavia, Italy}

\end{center}
\begin{abstract}

In a previous paper (hep-th/0306142) we have started to explore
the holographic principle in the case of asymptotically flat
space-times and analyzed in particular different aspects of the
Bondi-Metzner-Sachs (BMS) group, namely the asymptotic symmetry
group of any asymptotically flat space-time. We continue this
investigation in this paper. Having in mind a S-matrix approach
with future and past null infinity playing the role of holographic
screens on which the BMS group acts, we connect the IR sectors of
the gravitational field with the representation theory of the BMS
group. We analyze the (complicated) mapping between bulk and
boundary symmetries pointing out differences with respect to the
AdS/CFT set up. Finally we construct a BMS phase space and a free
hamiltonian for fields transforming w.r.t BMS representations. The
last step is supposed to be an explorative investigation of the
boundary data living on the degenerate null manifold at infinity.

\end{abstract}

\newpage

\tableofcontents
%%%%%%%%%%%%%%%%%%%%%%%%%%%%%%%%%%%%%%%%%%%%%%%%%%%%%%%%%%%%%%%%%%%%%%%%%%%%
%%%

%%%%%%%%%%%%%%%%%%%%%%%%%%%%%%%%%%%%%%%%%%%%%%%%%%%%%%%%%%%%%%%%%%%%%%%%%%%%
%%%

%%%%%%%%%%%%%%%%%%%%%%%%%%%%%%%%%%%%%%%%
\section{Introduction}
%%%%%%%%%%%%%%%%%%%%%%%%%%%%%%%%%%%%%%%%%%

The Holographic Principle gives a precise, general and
surprisingly strong limit on the information content of space-time
regions. Indeed, it seems to predict that in any theory containing
gravity, bulk degrees of freedom can be encoded holographically on
lower dimensional boundaries and arranged in such a way to give a
peculiar upper bound to the total number of independent quantum
states. The latter are indeed supposed to grow exponentially with
the surface
area rather than with the volume of the system.\\
In a previous paper \cite{Arcioni}, we have explored the
holographic principle in the context of asymptotically {\it flat}
space-times (notice that a different approach has been recently proposed in \cite{de
boer} and \cite{solodukhin}). We have considered quite in detail the asymptotic
symmetry group of {\it any} asymptotically flat space-time, namely
the Bondi-Metzner-Sachs (BMS) group. In particular we have derived
the covariant wave equations for fields carrying BMS
representations and made qualitative links between bulk and
boundary symmetries.\\
The purpose of this paper is to continue such investigations along
similar lines. The proposal is that holography in asymptotically
flat space-times is implemented via a S-matrix relating data
encoded on future and past null infinity and that the symmetry
group acting on such holographic screens is indeed the BMS group.
We then consider different aspects which (we believe) have to
taken into account in such a framework.\\
The paper is organized as follows: In Section 2 we give a brief
review of the BMS group and of its properties. For more details we
refer to our previous work \cite{Arcioni}, where we tried to make
a self contained introduction to the BMS group.\\
In Section 3 we explain in general terms the S-matrix approach we
follow. We underline similarities with 't Hooft \cite{gerard}
approach for the holographic description of black holes and in
particular with Ashtekar asymptotic quantization program
\cite{asymptoticquantization}. We point out at the same time the
difference with the AdS/CFT set up, the latter representing a
beautiful realization of holography in asymptotically Anti de
Sitter
space-times.\\
In Section 4 we consider the IR sectors of the gravitational field
and their link with the BMS group originally proposed by Ashtekar
(again in \cite{asymptoticquantization}). First we review the
issue of IR sectors in general and their standard re-summation via
Kinoshita-Lee-Neuenberg (KLN) theorem. We revisit the
Kulish-Faddeev approach \cite{kf} for QED, an alternative method
in which IR sectors are not summed and one ends up with
complicated asymptotic states. We then show how to generalize the
latter to the gravitational field and we make a connection between
the IR sectors of the gravitational
field and the BMS representation theory.\\
In Section 5 we use different results to analyze how bulk and
boundary symmetries are related in a complicated way. We show in
particular the strong dependence on the asymptotics of
space-time, in agreement with the holographic principle.\\
In Section 6 we recall a geometrical picture of the BMS group
using the natural fiber bundle structure of null infinity. This is
useful if one has in mind to construct a boundary theory with
fields carrying BMS indices.\\
Section 7 (more technical) represents a preliminary attempt to
construct a theory based on the BMS group. We therefore construct
a BMS phase space, a free BMS hamiltonian and analyze similarities
and differences with respect to the Poincar\'e case.\\
We finally end up with some concluding remarks and outlooks.\\

%%%%%%%%%%%%%%%%%%%%%%%%%%%%%%%%%%%%%%%%%%
\section{Brief review of the BMS group}
%%%%%%%%%%%%%%%%%%%%%%%%%%%%%%%%%%%%%%%%%%%
In this section we briefly review the BMS group and some of its
properties. We refer to our previous paper \cite{Arcioni} and
references therein for a detailed description.\\
The BMS group is the asymptotic symmetry group of {\it any}
asymptotically flat space-times. It can also be shown that it is
the isometry group of null infinity $\Im$ preserving its inner
degenerate metric and
the strong conformal geometry, the latter representing a generalized version
for a null hypersurface of the ordinary concept of angle.\\
In four dimensions\footnote{The BMS group changes according to the
dimension of bulk space-time because of the dimension dependent
fall-off properties of fields. In the whole paper we will always
consider {\it four} dimensional bulk space-times. In addition we
use the canonical definition \cite{penroserindler} of null
infinity $\Im$, assuming complete generators and $\Im \sim S^2
\times R$. In principle it is also possible to relax these
properties \cite{nullkilling}; the asymptotic symmetry group is in
this case different from the BMS in and may not even exists. All
these cases, however, turn out to be quite unphysical and in
general they do not exhibit gravitational radiation.}, where $\Im
\sim S^2 \times R$, the BMS is given by the semidirect product of
the (connected component of the homogeneous) Lorentz group with
the abelian group of real functions on the 2-sphere closed under
addition.\footnote{As suggested in \cite{Girardello} and
references therein, one has the freedom to choose a suitable
topology on the set of maps $f:S^2\to\Re$: one can impose a
"nuclear topology", i.e. $f\in C^\infty(S^2)$, or an Hilbert
topology i.e. $f\in L^2(S^2)$. From now on we consider the latter
and refer to \cite{Arcioni}, \cite{Mc4} for an analysis of the
differences between the two scenarios.} It is therefore similar to
the Poincar\'e group even if the translation subgroup is now
enlarged to the so called super-translations, i.e. functions on
the two sphere: expanding them into spherical harmonics we now
have an infinite number of coefficients entering in the expansion,
while in the case of translations we would have kept only four of
them. One therefore has an infinite dimensional group instead of
the ten dimensional Poincar\'e group. This is counterintuitive:
one would have expected the Poincar\'e group since gravity is
"weak" at infinity and this situation is also intriguing from the
point of view of the holographic principle, being quite different
with respect to asymptotically (A)dS space-times.\\
 Working with
the double cover of the Lorentz group (so as to get projective
representations as in standard quantum mechanics) the BMS group is
thus
$$BMS=ST\ltimes SL(2,\mathbb{C})$$
where $ST=L^2(S^2)$ are the super-translations. If we choose a
local chart $(u,\theta,\varphi)$ on $\Im$, any element lying in the BMS
group can be written as
$\left(\Lambda(\theta,\varphi),\alpha(\theta,\varphi)\right)$,
$\Lambda$ representing a conformal motion on the two sphere and
$\alpha$ a super-translation. Choosing then a point $x$ on the
2-sphere, the composition law between two group elements is
\begin{eqnarray}
(\alpha_1,\Lambda_1)(\alpha_2,\Lambda_2)=(\alpha_1+\Lambda_1\alpha_2,\Lambda
_1\Lambda_2),\\
\Lambda_1\alpha_2(x)=K_\Lambda(x)\alpha_2(\Lambda^{-1}x),
\end{eqnarray}
where $K_\Lambda(x)$ is a conformal factor (see \cite{Arcioni} for
a definition). There is no reference to the "u coordinate" in the
group transformation law. This reflects the peculiar nature of
$\Im$ which can be seen as a null sub-manifold of any
asymptotically flat space-time with degenerate metric
$$ds^2=0\cdot du^2+d\theta^2+\sin^2\theta d\varphi^2,$$
where "u" plays the role of an affine parameter (we will return to
this point in Section 6).\\
Recall that the BMS group contains a unique four-parameter
subgroup, the translations $T_4\subset ST$; however it is not
possible to extract a unique Poincar\'e subgroup from the BMS
group even if $ISO(3,1)\subset BMS$. On the other hand one has
$$g^{-1}ISO(3,1)g=ISO(3,1).$$
for any element $g\in BMS$ and in particular for any pure
super-translation (i.e. an element lying in $ST/T_4$). Thus there
are as many inequivalent Poincar\'e sub-groups contained in the
BMS group  as the number of elements of
$ST/T_4$.\\

%%%%%%%%%%%%%%%%%%%%%%%%%%%%%%%%%%%%%%%%%%%%%%%%%%%%%%%%%%%%%%%%%%%%%%%%%%%%
%%%%%%
\section{S-matrix approach for asymptotically flat space-times as a natural
arena
for the BMS group}
%%%%%%%%%%%%%%%%%%%%%%%%%%%%%%%%%%%%%%%%%%%%%%%%%%%%%%%%%%%%%%%%%%%%%%%%%%%%
%%%%%%

The most natural way to formulate holography in any asymptotically
flat space-time seems to be via a S-matrix mapping data collected
on past and future null infinity respectively. These "screens" can
be thought as abstract manifolds on their own and the S-matrix is
the operator mapping two Hilbert
spaces associated with past/future null infinity.\\
The construction of this operator is an enormously difficult task.
We assume, however, that such a map {\it exists}, satisfies some
natural requirements and analyze thus
the consequences.\\
This is similar to the strategy advocated by 't Hooft
\cite{gerard} with the S-matrix Ansatz for the physics of the near
horizon region of black holes, namely a map connecting data
encoded on the past and future horizons.\\
We will see first of all that the IR sectors of the gravitational
fields have to be considered with care when constructing such a
map. Most important, the natural symmetry group acting on the
screens $\Im^+,\Im^-$ is precisely the BMS group. We will try to
see if one can make connections between this group and bulk
symmetries and to understand better what kind of fields live on
the screens and are supposed to
transform with respect to BMS "indices".\\
A similar approach has been proposed by Ashtekar
\cite{asymptoticquantization} and goes under the name of
"asymptotic quantization", even if by that time the holographic
perspective was not yet known. The main idea in that case is to
promote to operators certain asymptotic quantities associated with
the gravitational field like the news tensor, the magnetic mass
and similar. If these represent the complete set of observables to
be quantized, this is still not clear. We nevertheless would like
to make a link between some of these results and the
representation theory of the BMS group, the latter having being
studied in great detail. We believe indeed that a better
understanding of the boundary symmetries and their link with the
bulk ones might be useful in a potential description of
holography. In addition, thinking in term of path integral, all
these observables should have a "weight" in the sum
over histories and therefore it is worth taking them into account.\\
There is of course a sharp difference with respect the well known
AdS/CFT correspondence. Banks \cite{tom} and others, in particular
Witten \cite{barioni}, have stressed (in a strong way) such
differences pointing out that the dual theory has to be something
non local or similar in the asymptotically flat case. We will
follow this point of view and as a consequence we would like to
see holography in asymptotically flat space-times {\it not} as the
flat space limit of AdS/CFT correspondence.

%%%%%%%%%%%%%%%%%%%%%%%%%%%%%%%%%%%%%%%%%%%%%%%%%%%%%%%
%%%%%%%%%%%%%%%%%%%%%%%%%%%%%%%%%%%%%%%%%%%%%%%%%%%
\section{BMS and gravitational IR sectors}
%%%%%%%%%%%%%%%%%%%%%%%%%%%%%%%%%%%%%%%%%%%%%%%%%%
The connection between IR sectors of the gravitational field and
the BMS group has been pointed out by Ashtekar several years ago
in \cite{asymptoticquantization}, where a detailed
analysis is given.\\
Here we simply recall the final result and then make a contact
with the representation theory of the BMS group, picking out the
BMS representations corresponding to the IR sectors of the gravitational
field.\\
Before doing this, however, we make a brief excursus reviewing
first the resummation of IR sectors normally performed in QED via
Kinoshita-Lee-Neuenberg (KLN) theorem and also the interesting
approach proposed by Kulish and Faddeev \cite{kf}, an alternative
method to get an IR-finite S-matrix for QED keeping this time into account
precisely the infrared sectors.\\
 We
show then how to generalize the latter to the gravitational case
to see explicitly the emergence of IR sectors
 and then move to the discussion of the IR sectors in the BMS framework.\\
%%%%%%%%%%%%%%%%%%%%%%%%%%%%%%%%%%%%%%%%%%%%%%%%%%%%%%%%%%%%%%%%%%%
\subsection{IR sectors in general and their common treatment via KLN
theorem}
%%%%%%%%%%%%%%%%%%%%%%%%%%%%%%%%%%%%%%%%%%%%%%%%%%%%%%%%%%%%%%%%
IR effects are normally associated with "soft" particles or
"collinear" jets of particles. The former are due to the fact that
any particle can be accompanied by an arbitrary number of massless
particles with vanishing momentum: such a collection will be
indistinguishable from the particle alone if the measured quantum
numbers are the same. Collinear jets, on the other hand, are due
to the fact that any massless particle can be indistinguishable
from an arbitrary number of massless particles with the same total
momentum and with the same sign of energy if their momenta are all
proportional since they all travel in the same direction at the
same speed.\\
Both these effects should then be taken into account in the
construction of the asymptotic states, though one normally
proceeds in a different way. Indeed, to any order in perturbation
theory only a {\it finite } number of potentially IR particles
contribute and this fact is the starting point of the well know
KLN theorem: to get IR finite cross sections one is instructed to
sum over both initial and final degenerate states which are
sources of potential IR effects. The S-matrix in turn will be IR
finite.  When summing over these degenerate states one has to
introduce a certain resolution in energy and angles, related to
the accuracy of the experimental apparatus. One tacit assumption
of the KLN theorem is that it is only the sum over initial/final
states which is experimentally measurable \cite{weinberg}.

%%%%%%%%%%%%%%%%%%%%%%%%%%%%%%%%%%%%%%%%%%%%%%%%%%%%%
\subsection{Asymptotic states and IR sectors: the QED example}
%%%%%%%%%%%%%%%%%%%%%%%%%%%%%%%%%%%%%%%%%%%%%%%%%%%%%%

An alternative and interesting way to get an IR finite S-matrix
has been proposed several years ago by Kulish and Faddeev
\cite{kf}. We just revisit the essential steps referring to the
original work for a detailed
description.\\
The main point is the following: one has a total hamiltonian
\begin{equation}
H = H_0 + H_{int}
\end{equation}
where $H_0$ is the free hamiltonian while $H_{int}$ is the
interaction piece which does not vanish at asymptotic times due to
the long range Coulomb potential describing the electromagnetic
field. The idea is then to consider a new hamiltonian where one
modifies the interaction part so that it vanishes at infinity. In
this case, however, also the free hamiltonian is modified and
becomes much more complicated. This is the "price" one has to pay
to get an IR free S-matrix.\\
One has therefore to define new fields and states and this is done
"dressing" the old ones by acting on them with a Moller operator.
The new fields $\phi'$, for instance, will be expressed in term of
the old ones $\phi$ as
\begin{equation}
\phi ' =U \phi U^{-1}
\end{equation}
and one can show that  U, the Moller operator, is
\begin{equation}
\label{hamiltonianasintotica} U(\tau)= T_- \exp \left( i \int^\tau
H^{asym}(\tau ') d\tau' \right)
\end{equation}
where $H^{asym}$ is the asymptotic hamiltonian, which is obtained
by considering the limit of the usual interaction hamiltonian (of
QED in this case) at large times.\footnote{$T_-$ is just a time
ordering operator which orders operators with time increasing from
left to right.} Note that the limit is not taken on the current
alone constructed from electron fields, but it is a limit
involving the photon field too. The different pieces entering into
the hamiltonian will contain terms like $e^{iat}$ and for large
times only those in which $a$ goes to zero will survive. Note that
one assumes small photon momentum in the computation and therefore
can set it to zero in all slowly momentum varying functions like
creation and annihilation operators. At the end of the day, the
photon field will be unchanged and one has an asymptotic current
coupled to it giving the asymptotic hamiltonian. The
asymptotic current is covariant by construction and has support on
the trajectory of a classical charge density moving uniformly with
respect to an "asymptotic" proper time.\\
After straightforward manipulations one can recast the Moller
operator in the form
\begin{equation}
U(\tau)= \exp  (-i \phi(\tau)) \exp(-R(\tau))
\end{equation}
The first term on the r.h.s is simply the relativistic Coulomb
phase factor associated with the long range Coulomb potential. The
second piece represents the radiative operator describing the
modification of the asymptotic states. One finds than that the
photon propagator is unchanged while the electron-positron one
changes and contains a so called distortion operator. The poles in
turn transform into a branch points and this is due to the photon
clouds surrounding the
charged particles.\\
One can show then that the radiative operator transforms the usual
Fock vector into a coherent state vector \cite{klauder} and the
corresponding coherent space is {\it not} unitarily equivalent to
a Fock space. At the end of the day these coherent states
represent the IR sectors of the electromagnetic field. The
Hilbert space is however still separable.\\
More in detail the radiative operator is of the form \be
\label{radiativeoperaotr} R=\exp \sum_i  \left( \alpha_i^* a_i
-\alpha_i a_i^\dagger \right) \ee i.e. the exponent of a linear
form of annihilation and creation operators of soft photons. It
can therefore be reduced to normal form \be R=
 \exp (-\frac{1}{2} \sum_i \mid \alpha_i \mid^2) \exp (- \sum_i \alpha_i
 a_i^\dagger) \exp(\sum_i \alpha_i^* a_i)
 \ee
If the series appearing in the first term on the r.h.s is finite
one has an ordinary Fock space. If not one has a coherent state
space. One can see that this is what happens in QED since the
series diverges in the IR limit and the divergence cannot be
eliminated by a renormalization procedure.

%%%%%%%%%%%%%%%%%%%%%%%%%%%%%%%%%%%%%%%%%%%%%%%%%%%%%%%%%%%%%%%%%%%%%%
\subsection{Emergence of coherent states in the case of gravity}
%%%%%%%%%%%%%%%%%%%%%%%%%%%%%%%%%%%%%%%%%%%%%%%%%%%%%%%%%%%%%%%%%%%%%%%%

We proceed in analogy with QED. We want to show that coherent
states describing soft gravitons appear as well when discussing
infrared singularities in the self energy of gravitons. We
therefore limit our analysis to the leading order approximation.
The derivation follows exactly the same steps of the original work
of Kulish and Faddeev \cite{kf}, once the appropriate
interaction hamiltonian has been chosen.\\
Consider then linearized gravity, i.e. take \be\label{weak}
 g_{\mu \nu} = \eta_{\mu \nu} +\gamma h_{\mu \nu}
 \ee
 with $h_{\mu \nu }$ describing the weak
gravitational field. We then plug the metric into the Einstein
action and expand up to trilinear order, since we need to take the
limit of an interaction hamiltonian. In this case we consider
graviton self-interactions. We truncate the expansion to the third
order since higher order terms would reduce the number of energy
denominators  which are responsible for the IR divergencies and
therefore also the degree of divergence \cite{weinberg}. Recall as
said that we simply want
to extract the leading order.\\
One then expands as usual the graviton field into annihilator and
creator operators \be
 \label{gravitonexp}
 h_{\mu \nu}=
\sum_{k,\lambda} \frac{1}{\sqrt{2 k_0 (2 \pi )^3}} \left(
\epsilon_{\mu \nu}(k,\lambda)\exp(-ikx)h_{k \lambda}+c.c. \right)
 \ee
and consider then the asymptotic large time limit of the
corresponding interaction hamiltonian. We can isolate the leading
IR part using previous experience with Kulish-Faddeev work. One
easily observes that in this case the leading terms will contain
two hard gravitons and one soft photons. One can then obtain a
drastic reduction of terms making a clever choice of gauge.
Normally one employs the so called De-Donder (harmonic) gauge.
However, using the similarity of GR with YM and remembering that
in the latter case \cite{siegel} the interaction hamiltonian
simplifies in the axial gauge. we use the same gauge also for
gravity. Plugging into the interaction hamiltonian the graviton
expansion one can further neglect - among the leading terms
containing as said one soft (with momentum $k_\mu$) and two hard
gravitons (with momenta $q^\mu$) - all terms proportional to soft
momenta but also those containing $\eta^{\mu\nu}\epsilon_{\mu\nu}$
and $q^{\mu} \epsilon_{\mu \nu}$ , these expressions vanishing
because of the physical condition on the polarization tensor. One
therefore is left with the simple expression \be
 H_{asympt}= \gamma \sum_{p,k} \sum_{\lambda,\sigma}
\frac{ K^{\mu \nu}(p,\sigma)}{\sqrt{2k_0 (2\pi)^3}} \left[
\epsilon_{\mu \nu}(k,\lambda) \exp (-i\frac{(p.k)t}{p_0}) h_{k
\lambda} +c.c. \right]
 \ee
with $\gamma$ defined in (\ref{weak}) and with
 \be K^{\mu \nu}= \frac{1}{p_0} p^\mu p^\nu h_{p \sigma}^\dagger h_{p
\sigma}
 \ee
Inserting the result in (\ref{hamiltonianasintotica}) one gets
after a straightforward manipulation an expression of the form \be
U= \exp (W-W^\dagger)
 \ee
where W is
\be
 W= \gamma \sum_{p,k} \sum_{\lambda,\sigma}
\frac{p_0}{(p.k)} \frac{K^{\mu \nu}(p,\sigma)}{\sqrt{2 k_0
(2\pi)^3}} \epsilon_{\mu \nu} (k,\lambda)  \exp
(-\frac{(p.k)t}{p_0}) h_{k\lambda} \ee W and its hermitian
conjugate are linear with respect to the annihilation and creation
operators of the soft gravitons. Recalling the discussion after
(\ref{radiativeoperaotr}) we therefore see that, again due to IR
problems, the soft gravitons live in a coherent vector space not
unitarily equivalent to a Fock space. We see the explicit
emergence of non trivial IR sectors in the case of the
gravitational field as well.\\
Now having in mind the holographic principle one would like to
have a characterization of the IR sectors at null infinity
starting however {\it directly} from the radiative degrees of
freedom at infinity. But this is precisely the framework proposed
by Ashtekar \cite{asymptoticquantization} where the BMS plays then
a crucial role in the universal classification of the IR sectors.
Note also that in this way one avoids the Fourier transformation
to annihilator and creator operators which would depend on the
details of the
asymptotics of space-time.\\
Before moving to the description of IR sectors via BMS, we would
like to point out the sharp difference with respect to
asymptotically Anti de Sitter (AdS) space-times. In the latter
case the boundary is time-like and one has the vanishing of
gravitational Bondi news \cite{magnon}, which as we will see below
are related to the IR sectors of the gravitational radiation. This
has remarkable consequences in terms of the definition of the
Cauchy problem ensuring the uniqueness of the solution
\cite{friedrich}. In the case of null infinity, on the other hand,
the non vanishing Bondi news brings about geodesic deviation even
for an arbitrary small energy flux \cite{ludvigsen}. Radiation at
infinity is therefore a cooperative phenomenon which cannot be
traced to the individual particle fields and as we will see the
rich IR structure is embedded in a complicate way in the
BMS group representations.\\

%%%%%%%%%%%%%%%%%%%%%%%%%%%%%%%%%%%%%%%%%%%%%%%%%%%%%%%%
\subsection{Gravitational field IR sectors and their BMS description}
%%%%%%%%%%%%%%%%%%%%%%%%%%%%%%%%%%%%%%%%%%%%%%%%%%%%%%%%

According to Ashtekar analysis the IR sectors of the gravitational
field are labelled by the quantity \be Q_{\mu \nu }(\theta,\phi)=
\int_\Im du N_{\mu \nu} (u,\theta,\phi)
 \ee
where $N_{\mu \nu}$ is \footnote{Here $\mu ,\nu =(u, \theta,
\phi)$} the news tensor, a sort of covariant version of the news
function of Sachs et. al. measuring the amount of radiation at
null infinity (see below for a more precise connection with the
radiative phase space
of the gravitational field).\\
In particular when $Q_{\mu \nu}=0$ then one has a trivial IR
sector and a Fock space description of the soft gravitons; when it
is a non vanishing, one has a coherent state of soft gravitons,
mathematically described by representations {\it} not Fock
equivalent (realized via Ge'lfand-Nairmark-Segal (GNS)
construction in more formal terms). These are the non trivial
sectors of the gravitational field.\\
$Q_{\mu \nu}$ is thus isomorphic to ST/T, i.e. to purely
super-translations. Before commenting on this,  we observe that
from physical point of view this means that $Q_{\mu \nu}$ tells us
how much good cuts are super-translated due to gravitational
radiation. In other words $Q_{\mu \nu}$ represents the obstruction
to reduce to Poincar\'e  group and can be then thought as a sort
of internal charge, just like in 1+1 massless QFT models  where
one has indeed additional non Fock representations associated with
a "topological" charge describing non trivial IR
sectors \cite{frohlich}.\\
We consider therefore in more detail the link between the "charge"
$Q_{\mu \nu }$ and ST/T, i.e. purely super-translations of the BMS
group.\\
Recall that the news tensor is always given by $N_{\mu \nu} = -2
\mathcal{L}_n  \gamma_{\mu \nu} $ where $\gamma_{\mu \nu}$
represent the radiative modes of the gravitational field at null
infinity \cite{asymptoticquantization} and $\mathcal{L}_n$ is the
Lie derivative w.r.t the normal of $\Im$. It can be shown in turn
that $\gamma_{\mu \nu}$ expresses the difference of connections
${D}$ (the news tensor is just the field strength) defined at null
infinity. Now the radiative modes belong to an ordinary Fock
graviton space if the inner product defined on the corresponding
phase space of radiative modes is finite. This conditions are
shown to be realized precisely when $Q_{\mu \nu}=0$ and in terms
of radiative modes when $\delta \gamma= \gamma_{\mu \nu}
(u=\infty) - \gamma_{\mu \nu} (u=- \infty)=0$, i.e. same
connections $D^0$ describing a classical vacuum both at $u= \infty
$ and $u=-\infty$. However, one can show that under a translation
the classical vacuum is invariant, while one can "jump" from one
vacuum to another precisely by mean of a super-translation.
Therefore when $\delta \gamma$ is non vanishing, i.e. in turn
$Q_{\mu \nu}$ is non zero, one has moved from one classical vacuum
to another via a super-translation and this is due to the presence
of gravitational radiation as said
many times.\\
Note that the radiative modes of the gravitational field
$\gamma_{\mu \nu}$ are {\it exact} solutions of Einstein equations
as one approaches null infinity and naturally sees the BMS group.
In the conventional treatment, on the other hand, one considers
small fluctuations $h_{\mu \nu}$ around flat space-time and the
covariance is with respect to the {\it latter}, i.e. the usual
Poincar\'e group.\\
We now make a connection with the BMS representation theory. We
have to pick out the purely super-translational representations in
ST/T. The latter is an infinite dimensional and not normal
subgroup contrary to the BMS translation subgroup . Let
$I=SO(3,1)\ltimes \frac{ST}{T^4} $. A direct study of the theory
of representations in \cite{Mc1}, \cite{Mc2} and of the wave
equations in \cite{Arcioni} shows that the spin degrees of freedom
labelling a BMS field are not necessarily related to I but they
are related to the representations induced from the BMS little
groups; moreover in \cite{Mc2} it was shown that the theory of
representations for the group I can be directly derived from the
equivalent theory for the full BMS group; in particular a
representation of I is equivalent to an unfaithful representation
of the BMS group and thus it is induced as well from a little
group through Mackey theory. Since the abelian subgroup of I is
equivalent to $ST$, we can easily see that all the little groups
of BMS are as well little group of I with the only exception of
$SU(2)$.\\
It is straightforward to extend the above considerations from the
BMS group to the I group thus concluding that all the BMS little
groups are as well I little groups with the only exception of
$SU(2)$ since it would have a vanishing fixed point. All these
representations are unfaithful.\\
Note that the group I was originally proposed by Komar
\cite{komar} to describe internal symmetries of elementary
particles- a perspective which was then abandoned. We consider
again the same group which stands indeed for an internal charge,
but labels the IR sectors of the gravitational field.\\
 In particular notice that
they turn out to have a rich and complicated structure. We find
intriguing the presence of non-connected little groups, telling us
that the gravitational field at infinity in these cases exhibits
somehow a sort of "christallographic" structures because of these
super-translations associated with non connected groups. The
presence of this mixture of spins from BMS reps theory receives
also now a better clarification, since they are associated to
configurations of soft gravitons appearing as a coherent
superposition of multi spins, the latter concept being somewhat
difficult to accept when referred as by Komar to
elementary particles.\\

%%%%%%%%%%%%%%%%%%%%%%%%%%%%%%%%%%%%%%%%%%%%%%%%%%%%%%%%
\section{Bulk-boundary symmetries}
%%%%%%%%%%%%%%%%%%%%%%%%%%%%%%%%%%%%%%%%%%%%%%%%%%%%%%%%

A key aspect of AdS/CFT  correspondence is the matching between
bulk and boundary symmetries. In the case of asymptotically flat
space-times the situation is definitively more complicated. We
give here a qualitative picture pointing out in particular the
role of the gravitational radiation at null infinity, which makes
enormously difficult to define a precise
mapping between bulk and boundary symmetries.\\
Before proceeding note that using the freedom in the choice of
screen one could as well select $i_0$, i.e. spatial infinity. Here
one has again an infinite dimensional asymptotic symmetry group,
the so called Spi group (see \cite{hansen} and \cite{Mc7}) and remarkably one can
pick out a {\it canonical} Poincar\'e subgroup group choosing
suitable boundary conditions. $i_0$ is however just a point
attached to the Penrose diagram  and not the natural boundary of
spacetime as $\Im$; in addition it captures the Coulomb part of
bulk fields, not the radiative aspects we are interested in.\\
Consider then the bulk isometry algebra $B$; it can be decomposed
\cite{xan},\cite{nullkilling} into the direct sum of $\tau$ plus
$B/ \tau$, with $\tau$ representing bulk Killing vectors fields
which give BMS super-translations when evaluated on $\Im$. When
$K_{abcd}n^d$ is zero \footnote{Recall $K_{abcd}= \Omega^{-1}
C_{abcd}$ with $\Omega$ the conformal factor, $C_{abcd}$ the Weyl
tensor and $n^a$ the normal on $\Im$.} on $\Im$ these turn out to
be BMS translations. Therefore BMS super-translations are not
originated from bulk symmetries in this case. However, this
condition is too strong in general and when it is not imposed bulk
symmetries {\it can} give rise effectively to boundary BMS
super-translations.\\
 Comparing for instance  with $AdS_3/CFT_2$ correspondence
we see similarity but also great difference: in that case one also
has an infinite dimensional conformal group on the two dimensional
boundary (once suitable b.c. are chosen) and apparently there is a
mismatch with the $SL(2,R) \times SL(2,R)$ $AdS_3$ bulk isometry
group. However only the the generators $L_{-1},L_{0},L_{1} $ of
the corresponding Virasoro algebra annihilate the vacuum and the
others, {\it not} arising from physical space-time symmetries, are
supposed to describe excited states, therefore are spectrum
generating. Similar considerations apply for AdS/CFT on coset
spaces \cite{strominger},\cite{marika} with degenerate boundaries.\\
In the case of asymptotically flat space-times, however, bulk
symmetries can give rise to super-translations. More precisely one
can actually show that (in the un-physical space-time)
 \be \nabla_{(a}\xi_{b)}= K g_{ab} + \Omega
X_{ab}. \ee On the boundary the last term on the r.h.s. clearly
vanishes so the quantity $X_{ab}$ can be interpreted  as the
failure of the BMS generator to arise from a bulk symmetry. For a
translation one has \be
  X_{ab}-\frac{1}{2}Xh_{ab} =\alpha N_{ab},
\ee while for a super-translation \be X_{ab}=- \frac{1}{2}\alpha
\beta g_{ab} + \sigma g_{ab} - \frac{1}{2}\alpha(R_{ab}-
\frac{1}{6}R g_{ab}) - \nabla_a \nabla_b \alpha, \ee with \be
 \Omega^2 \beta= n^a n_a  \ee and
 \be \Omega \sigma= n^a \nabla_a
\alpha.
\ee
Notice that the vanishing of $X_{ab}$ depends in a non
trivial way from the size $\alpha$ of the BMS group. Recall that
"time" translations are generated by $\alpha (\theta,\phi)
\partial /
\partial u$ the size being arbitrary on each fiber. So the time
evolutions is related with the issue of bulk-boundary symmetries
in a
complicated way.\\
In addition observe that the quantities $\beta$ and $\sigma$
contain information about the generator $n^a$ and the conformal
factor $\Omega$: the percolation of light rays close to the
boundary is therefore related again to the issue of symmetry
mapping.\\
We see how crucial is the asymptotic region of space-time in the
holographic principle and
how complicated is in the case of asymptotically flat space-times.\\
One can further show that $B/ \tau$ corresponds to a Lorentz
subalgebra acting on the $S^2$ spheres sections of the null
boundary. Some of these Lorentz subalgebras {\it are} indeed the
little groups of the BMS representation theory but some not
therefore making difficult to realize an intriguing bulk-boundary
matching.\\

%%%%%%%%%%%%%%%%%%%%%%%%%%%%%%%%%%%%%%%%%%%%%%%%%%%%%%%%%%%
\section{Geometrical description of $\Im$ and of its "Fields"}
%%%%%%%%%%%%%%%%%%%%%%%%%%%%%%%%%%%%%%%%%%%%%%%%%%%%%%%%%%%%%%%
In a S-matrix approach with data stored at past and future null
infinity covariance on these two "screens" is expressed in terms
of the BMS group, which characterizes intrinsically $\Im$ and its
universal geometry. An elegant, geometrical and coordinate
independent description of $\Im$ appears in \cite{sommers}. This
helps us to understand the
notion of BMS fields and eventually of the "BMS hamiltonian" constructed out
of them.\\
According to this geometrical interpretation, once $\Im$ is given
a natural fiber bundle structure, with base space $S^2$, a unique
and universal characterization of the BMS group is obtained.
Namely, a BMS symmetry transformation is simply the lift of the
conformal motions on $S^2$ to $\Im \sim R \times S^2$, with the
requirement that a (2,2) tensor S on $\Im$ is preserved\footnote{Using coordinates one has $S^{ab}_{cd}=q_{cd}n^a n^b$
with $q_{ab}$ induced metric on $\Im$ and $n^a$ the normal,
precisely the same tensor which enters in \cite{geroch} in the
definition of universal geometry of $\Im$.}. This restricts the
lift and the function $\alpha ( \theta,
\phi)$ -the "size" of the BMS group- is the only remnant of the freedom in
the lifting.\\
Given thus a conformal motion on the sphere one lifts it to $\Im$:
at fixed angles one moves along R, i.e. "the time", (the size
$\alpha$ being different on each fiber!) while changing the angles
one will jump to another fiber. On this fiber bundle structure one
defines a (p,q)
tensor field which is then a field carrying BMS indices by construction.\\
We would like to observe that $u$ is an affine parameter not a
coordinate (reflecting the degeneracy of the $\Im$) and  one does
not really have a time evolution. This further supports a S-matrix
description mapping two Hilbert spaces associated with past and
future null infinity respectively, which are supposed to be
"kinematical" the dynamics being given by the S-matrix. The data
are collected on the two dimensional $S^2 $ sections and moving
along u is just a way to label them.
%%%%%%%%%%%%%%%%%%%%%%%%%%%%%%%%%%%%%%%%%%%%%%%%%%%%%%%%%%%%%%%%%%%%%%%%
\section{BMS Phase space, hamiltonian and Poincar\'e reductions}
%%%%%%%%%%%%%%%%%%%%%%%%%%%%%%%%%%%%%%%%%%%%%%%%%%%%%%%%%%%%%%%%%%%%%%%%%

As we have pointed out many times in the previous Sections, the
symmetry group which naturally lives on $\Im$ is the BMS group.
Boundary fields should therefore carry indices w.r.t such a group.
In this section we begin to construct such fields and their free
hamiltonian, making comparison with the Poincar\'e case at the
end.\\
This section is quite technical. The main point, however, is that
what is quite unconventional is not the fact that we have to do
with an infinite dimensional group (after all this is quite common
in string theory or similar approaches since gravity introduces
many degrees of freedom \cite{igor}) but the issue of the time
evolution. Roughly this means moving along $u,v$ coordinates on
future and past infinity respectively. It is here that the
degenerate nature of $\Im$ enters into the game. Despite the fact
that there is some additional structure on $\Im$, namely the
strong conformal geometry preserved by the BMS, this is quite
"poor" compared with usual non degenerate metrics.\\
The super-translation enlargement, when looked from {\it boundary}
point of view, is a consequence of this lack of structure. This is
responsible for the various departures from conventional QFT as
pointed out below.\\
%%%%%%%%%%%%%%%%%%%%%%%%%%%%%%%%%%%%%%%%%%%%%%%%%%%%%%%%%%%%%%%%%%%%%%%%%%%%
%%%%%%%%%
\subsection{Canonical and covariant phase space}
%%%%%%%%%%%%%%%%%%%%%%%%%%%%%%%%%%%%%%%%%%%%%%%%%%%%%%%%%%%%%%%%%%%%%%%%%%%%
%%%%%%%%%%%
According to the previous considerations one should proceed in a
coordinate independent way to construct the BMS phase space. The
prescription (of great generality) which gives the so called
covariant phase space is described in the elegant works
\cite{Witten},\cite{Crnkovic} which
we follow for our purposes.\\
In order to describe the key concepts, let us consider first  for simplicity
a scalar field
and recall that the \emph{canonical} phase space is given by
$$\Gamma=\left\{\varphi,\pi\right\}.$$
where $\varphi=\phi\mid_{\Sigma_3}$ is the value of the field on a
suitable three dimensional space-like hyper-surface and $\pi$ is
the conjugate momentum. This is normally realized via a 3+1
splitting of space-time in the usual ADM like framework. In the BMS
invariant system that we consider such approach is difficult to apply
since it is not
evident how to single out a "time direction" due to the degeneracy of
null infinity.\\
Nonetheless there is another way to deal with a phase space which
goes under the name of \emph{covariant} phase space. The latter is
the set of solutions (up to some regularity conditions) of the
classical equation of motions\footnote{Notice that in an
Hamiltonian framework, one identifies the solutions of the
equation of motions as integral curves of the hamiltonian itself.
This interpretation does no longer apply to the covariant phase
space where from one side it is still possible to study the time
development of a system through its hamiltonian vector field
whereas from the other, the hamiltonian flow takes us from one
solution (i.e. a point) in the phase space to a different solution
infinitesimally close. Within this approach the study and the
properties of physical system goes as in the standard picture.}.
For a scalar field, one has
$$\Gamma_s=\left\{\phi\in C^\infty(M^4)\;|\;
\nabla^a\nabla_a\phi-m^2\phi=0\right\}.$$ Let us stress that in
this special case, there is a full equivalence between $\Gamma_s$
and $\Gamma$ which can be proved observing in turn that there is a
one to one correspondence between the set of solution of
Klein-Gordon equation and the set of initial data chosen on a
generic $\Sigma_3$ hyper-surface \cite{Corichi}. In a general
framework instead, the covariant phase space $\Gamma_s$ is a
subset of $\Gamma$ and whereas the first takes into account all
the kinematically possible configurations, the latter describes
the {\it dynamically possible configurations} for the fields.\\
A key difference between the canonical and the covariant phase
space that we wish to emphasize is that, whereas the former breaks
the symmetries of the system, the latter by construction preserve
it. Furthermore, let us emphasize that, in the framework of
general relativity and more generally in diffeomorphism covariant
lagrangian field theories, the covariant phase space is the
natural playground where to analyze the general relationship
between local symmetries and constraints  and it has allowed for a
full definition of conserved quantities in particular on future
and past null infinity \cite{Lee},\cite{Iyer},\cite{Wald}.

%%%%%%%%%%%%%%%%%%%%%%%%%%%%%%%%%%%%%%%%%%%%%%%%%%%%%%%%%%%%%%%%%%%%%%%%%%%%
%%%%%
\subsection{BMS phase space}
%%%%%%%%%%%%%%%%%%%%%%%%%%%%%%%%%%%%%%%%%%%%%%%%%%%%%%%%%%%%%%%%%%%%%%%%%%%%
%%%%%%%%%

To construct a phase space for a BMS field we need to consider BMS
covariant wave equations. These have been derived in our previous
paper \cite{Arcioni} and discussed in great detail.\\
Relying then on previous considerations, one has that the BMS
invariant phase space has to be
$$\Gamma_{BMS}=\left\{\phi:N=L^2(S^2)\to\mathbb{C}\;|\;\phi\in
L^2(\mathcal{O})\otimes\mathcal{H}^\lambda\right.$$
\begin{equation}
\label{BMSphase}
\left.{\rm
with}\;\pi(\alpha)\phi(\alpha)=\phi(\alpha)\bigg\}\right.,
\end{equation}
where $\mathcal{O}$ is an orbit  for one of the little groups and
$\mathcal{H}^\lambda$ is an Hilbert space isomorphic to
$\mathbb{C}^{(2j_1+1)(2j_2+1)}$ carrying an $SL(2,\mathbb{C})$
label $\lambda=(j_1,j_2)$ . Recall \cite{Arcioni} that the space
$N$ is the super-momentum space, i.e. the BMS equivalent of the
space of $p^\mu$ vectors of the Poincar\'e case. Furthermore let
us stress that the equations of motion are two-folds: from one
side we have the support condition on the orbit (i.e. $\phi\in
L^2(\mathcal{O})$) for the wave equation which is the equivalent
in the BMS case to the Klein-Gordon equation for a Poincar\'e
field while the other represents the ortho-projection
$\pi(\alpha)\phi(\alpha)=\phi(\alpha)$ which picks out some
components in $\mathbb{C}^{(2j_1+1)(2j_2+1)}$, i.e. a subspace
$\mathbb{C}^m$, carrying a representation of the corresponding
little group. Notice also that a nice feature of the BMS group is
that all the little groups are compact and thus the structure of
(\ref{BMSphase}) is in some sense universal. On the other hand, in
the Poincar\'e case, massless field live on the orbit
$SL(2,\mathbb{C})/E(2)$ and carry unfaithful(!) representations of
the non compact little group $E(2)$,\footnote[3]{Technically
speaking, the situation for massless fields is rather complicated
and a correct understanding of the symplectic phase space and of
the physical degrees of freedom requires to mod out gauge degrees
of freedom
via Marsden-Weinstein reduction theorem \cite{bookL}.}\\
Furthermore, notice that in (\ref{BMSphase}) the covariant wave equations
are first class
constraints so that the covariant phase space is a linear complex
Hilbert space; therefore, one can introduce a canonical strongly non
degenerate
symplectic two form on the infinite dimensional manifold $\Gamma_{BMS}$
\cite{book2}:
\begin{eqnarray}
\Omega:\Gamma\times \Gamma\to\Re\\
\Omega(\phi,\varphi)=Im\int\limits_{\mathcal{O}} d\tau
\left(\bar{\phi}^\mu\varphi_\mu\right),\label{BMSsymp}
\end{eqnarray}
where $d\tau$ is the natural measure induced on $\mathcal{O}$  (for the
BMS orbits see \cite{Mc1}, \cite{Mc2}) and $\mu$ is an index
labelling the component of the function $\phi$ in
$\mathcal{H}^\lambda$ (for an analogue construction in the Poincar\'e
context see \cite{Landsman}) .\\
At the end of the day we are dealing with a covariant phase space
(\ref{BMSphase}) endowed with the above symplectic form. We can now
proceed with the construction of the corresponding (free)
hamiltonian.
%%%%%%%%%%%%%%%%%%%%%%%%%%%%%%%%%%%%%%%%%%%%%%%%%%%
\subsection{BMS free Hamiltonian}
%%%%%%%%%%%%%%%%%%%%%%%%%%%%%%%%%%%%%%%%%%%%%%%%
We follow Chernoff and Marsden \cite{Chernoff} (see also \cite{streubel}). In particular
recall that, whenever we deal with a phase space $\mathcal{E}$
with the structure of a complex Hilbert space endowed with a
symplectic form $\Omega$ invariant under the action of a one
parameter group $G$ with generator $A$, we can associate to a
field $\phi\in\mathcal{E}$ an energy function:
\begin{equation}
H(\phi)=-\frac{1}{2}\Omega(iA\phi,\phi)\label{hamiltonian}
\end{equation}
In particular, if we consider a phase space invariant under the action of a
Lie
group $G$ with Lie algebra $\mathfrak{g}$ whose generators satisfy the usual
commutation relation
$$\left[\zeta_a,\zeta_b\right]=f_{abc}\zeta^c,$$
it possible to associate to each generator $\zeta_a$ the one
parameter group $U_t(\zeta_a)=e^{t\zeta_a}$ and $\zeta_a$ acts on
the elements of the phase space through the Lie derivative
$\mathcal{L}_{\zeta_a}$. Bearing in mind the above remarks and the
natural connection between the induced hamiltonian action and the
Mackey construction of induced representations \cite{Zakrzewski},
we can apply the construction to the fields associated to the BMS
group whose Lie algebra is known \cite{sachs}:
$$\mathfrak{bms}=\mathfrak{sl}(2,\mathbb{C})\oplus L^2(S^2).$$
Moreover let us stress that the exponentiation of this infinite dimensional
Hilbert-Lie
algebra is strictly related to the global properties of $\Im$, in
particular to the completeness of the geodesic generators of the
boundary which grant us that $\Im$ is globally and not only
locally $S^2\times\Re$.
Eventually for any BMS field, we can define through (\ref{BMSsymp}) an
Hamiltonian function associated
to each generator $\zeta_a$:
\begin{equation}\label{energy}
H_{\zeta_a}(\phi)=-\frac{1}{2}\Omega(i\mathcal{L}_{\zeta_a}\phi,\phi)=\frac{
1}{2}Re\int\limits_{\mathcal{O}}d\tau\left(\overline{\mathcal{L}_{\zeta_a}\phi}(p)\phi(p)\right).
\end{equation}
In the next Section we comment further on the
above energy function, more specifically on its relation with the
Poincar\'e energy functions.
%%%%%%%%%%%%%%%%%%%%%%%%%%%%%%%%%%%%%%%%%
\subsection{BMS vs Poincar\'e}
%%%%%%%%%%%%%%%%%%%%%%%%%%%%%%%%%%%%%%%%%%%%%%%%%

We point out in the following differences and similarities between
BMS and Poincar\'e group. We use the results of \cite{Arcioni} and
of the previous sub-sections focusing in particular on the
massive and massless representations. We choose the little group
$SU(2)$ describing massive particles and then on $SO(2)$
describing massless particles. Similar considerations extend to
the other little groups both in the massive and massless case.\\
\begin{center}
{\it $SU(2)$ massive case}
\end{center}
The associated wave equations live on an orbit isomorphic to
$\Re^3$ and they transform under a representation of the full
Lorentz group whereas the equation of motions are an Hilbert space
reduction. Taking into account (\ref{BMSphase}), the covariant phase space
associated with $SU(2)$ respectively in the BMS and in the Poincar\'e case
is:
$$\Gamma_{BMS}=\left\{\psi:\frac{SL(2,\mathbb{C})\ltimes N}{SU(2)\ltimes
N}\sim
\frac{SL(2,\mathbb{C})}{SU(2)}\to\mathbb{C}\;\;\;|\;\;\;\psi\in
L^2(\Re^3)\otimes\mathcal{H}^\lambda\;\;\;\right.$$
\begin{equation}
\left.\pi(\alpha)\psi(\alpha)=\psi(\alpha)\right.\bigg\}\label{ph1}
\end{equation}
$$\Gamma_{P}=\left\{\psi:\frac{SL(2,\mathbb{C})\ltimes T^4}{SU(2)\ltimes
T^4}\sim
\frac{SL(2,\mathbb{C})}{SU(2)}\to\mathbb{C}\;\;\;|\;\;\;\psi\in
L^2(\Re^3)\otimes\mathcal{H}^\lambda\right.$$
\begin{equation}
\left.\;\;\;\pi(p)\psi(p)=\psi(p)\bigg\}\right.\label{ph2}
\end{equation}
One has that the homogeneous space where the particle live is the
same in both cases and also the equations of motion are related:
if we reduce performing a Lorentz transformation the projection
equation from a generic value of respectively $\phi$ and $p$ to
the fixed point on the orbit (i.e. the center of mass), one has
(\cite{Arcioni},\cite{Barut}) respectively
$\pi\psi(\bar{p})=\psi(\bar{p})$ and
$\pi\psi(\bar{\phi})=\psi(\bar{\phi})$ where $\bar{p}$ is the four
momentum $p^\mu=(m,0,0,0)$ in $T^4$ and $\bar{\phi}=m$ is the
constant super-translation. Thus in the BMS case there is a
vanishing pure super-translational part.\\
Interestingly the BMS fixed point is
completely equivalent to the Poincar\'e fixed point.\\
Despite the similarity, however, differences appear when we
consider the hamiltonian for the massive particle because the
symplectic form associated to the phase space (\ref{ph1}) is
invariant under the Poincar\'e group whereas the symplectic form
associated to (\ref{ph2}) is invariant under the full BMS group.
This difference is thus reflected in the construction of the
hamiltonian which depends on the symplectic form and on the
generators of the symmetry algebra.
Actually one has for BMS and Poincar\'e respectively
$$H_{\zeta_a}(\phi)=\frac{-1}{2}\Omega(i\mathcal{L}_{\zeta_a}\phi,\phi)$$
\begin{equation}
=\frac{-1}{2}
Im\int\limits_{\Re^3}d\mu
[-i\overline{\mathcal{L}_{\zeta_a}\phi}^\lambda(x)]\phi_\lambda(x),\;\;\;
a=1,...,\infty
\end{equation}
$$H_{\zeta_a}(\phi)=\frac{-1}{2}\Omega(i\mathcal{L}_{\zeta_a}\phi,\phi)=$$
\begin{equation}
\frac{-1}{2}
Im\int\limits_{\Re^3}d\mu
[-i\overline{\mathcal{L}_{\zeta_a}\phi}^\lambda(x)]\phi_\lambda(x),\;\;\;a=1
,...,10
\end{equation}
where $x$ is a point over the orbit and where in the first
expression we have an infinite number of possible Hamiltonian
functions due to super-translations
whereas in the Poincar\'e scenario the number is finite .\\
Recall as a matter of fact that the vector valued fields live on
an homogeneous manifold $\frac{G}{H}$ whereas the Hamiltonian is
constructed with the generators of the Lie algebra of the full $G$
group. Let us thus choose a basis $\left\{\zeta_a\right\}$ for the
Lie algebra of $G$ and let us extract a subset which is a basis
for the algebra of $H$. Let us pick one of such generators,
$\xi_a$ and let us apply it to a field living on $\frac{G}{H}$:
$$\mathcal{L}_{\xi_a}\phi(x)=\frac{d}{d\lambda}(\exp^*(\lambda\xi_a)\phi(x))
|_{\lambda=0}=
\frac{d}{d\lambda}\phi(\exp(\lambda\xi_a)x)|_{\lambda=0}=$$
$$=\frac{d}{d\lambda}\phi(hx)|_{\lambda=0}
=\frac{d}{d\lambda}\phi(x)|_{\lambda=0}=0,$$ where in the first
two equalities we use the definition of Lie derivative and
exponential map, $h$ is an element in $H$  and $hx=x$ since $x$
lies in the coset space $\frac{G}{H}$. This translates in our
scenario in the cancellation in the hamiltonian of all the
contributes respectively from $SU(2)\ltimes T^4$ for the
Poincar\'e group and from $SU(2)\ltimes N$ for the BMS group.
Considering also the previous remarks on the fixed point for the
BMS group, the possible hamiltonians are identical in both cases:
\begin{equation}\label{massiveh}
H_{\zeta_a}(\phi)=\frac{-1}{2}Re\int\limits_{\mathcal{O}\sim\Re^3}\overline{
\mathcal{L}_{\zeta_a}\phi}(p)\phi(p)
d\mu(\mathcal{O}).\;\;\;a=1,...,dim\frac{SL(2,\mathbb{C})}{SU(2)}
\end{equation}
%%%%%%%%%%%%%%%%%%%%%%%%%%%%%%%%%%%%%%%%%%
\begin{center}
{\it $SO(2)$ massless case}
\end{center}
%%%%%%%%%%%%%%%%%%%%%%%%%%%%%%%%%%%%%%%%%%%
In the BMS group, massless fields are associated with $\Gamma$ the double
cover of the $SO(2)$ little group whereas in the Poincar\'e case
we deal with the non compact $E(2)$ group.\\
In the BMS case the situation is therefore rather similar to the
$SU(2)$ little group. Massless particles live on the orbit
\begin{equation}\label{covmassl}
\psi:\frac{SL(2,\mathbb{C})}{\Gamma}\sim\Re^3\times S^2\to \mathcal{H}^s,
\end{equation}
where
$\mathcal{H}^s=L^2(\frac{SL(2,\mathbb{C})}{\Gamma})\otimes\mathcal{H}_s$.
These wave functions transform under a representation of the full
$SL(2,\mathbb{C})$ group and the equations of motions reduce the
carrier space $\mathcal{H}_s$ to a sub-Hilbert space carrying
representations of the $SO(2)$ little group which is
one-dimensional. The equations of motion can be represented as the
usual ortho-projection:
$$\pi(\phi)\psi(\phi)=\psi(\phi),$$
where $\phi$ is a point in $L^2(S^2)$ lying on the orbit
$\mathcal{O}=\frac{SL(2,\mathbb{C})}{\Gamma}$. Massless particles
can then  be considered as the covariant wave equations
(\ref{covmassl}) satisfying the above equations of motions
(for details we refer to \cite{Arcioni}, \cite{Asorey}).\\
Considering as usual the covariant description, the phase space
for BMS massless particles (with positive energy) becomes:
$$\Gamma_{BMS}^{m=0,+}=\left\{\psi:\frac{SL(2,\mathbb{C})}{\Gamma}\sim\Re^3\times
S^2\to \mathbb{C}\;\;|\;\;\psi\in
L^2(\frac{SL(2,\mathbb{C})}{\Gamma})\otimes\mathcal{H}_\lambda\right.$$
\begin{equation}
\label{phasemassl}
\left. {\rm and}\;\;\pi(\phi)\psi(\phi)=\psi(\phi)\bigg\}\right.
\end{equation}
The covariant phase space for massless particles is a symplectic space with
a
symplectic form identical to (\ref{BMSsymp}) and thus, following the same
construction as in the massive case, we can associate to each Lie algebra
symmetry generator an hamiltonian:
\begin{equation}\label{hammassl}
H_{\zeta_a}(\psi)=\frac{-1}{2}\Omega(i\mathcal{L}_{\zeta_a}\psi,\psi)=
\frac{1}{2}Re\int\limits_{\Re^3\times
S^2}d\mu\;\overline{\mathcal{L}_{\zeta_a}\psi}(p)\psi(p),
\end{equation}
where $\zeta_a\in\frac{\mathfrak{sl(2,\mathbb{C})}}{\mathfrak{so(2)}}$ and
$d\mu$ is the
natural measure on $\mathcal{O}$ (see the appendix of
\cite{Mc1}).\\
This hamiltonian is rather similar to (\ref{massiveh}) since it is
constructed in a similar way but there is a main difference since
the wave equations live now on a five dimensional manifold.\\
This is in sharp contrast with massless particles transforming under a
Poncar\'e representation and living in a three dimensional variety; the main
reason for this difference is related to the infinite dimensional
nature of the BMS group.\\
Actually the homogeneous manifold $\frac{SL(2,\mathbb{C})}{SO(2)}$
is constructed through the action of the full BMS group on a point
$p=p(\theta,\varphi)$ in $L^2(S^2)$ fixed under the action of the
isotropy group $\Gamma$. As shown in $\cite{Mc2}$ the point $p$ has
the form
\begin{equation}\label{fixedp}
p(\theta,\varphi)=p_0+p_3\cos\theta+\alpha(\theta),
\end{equation}
where $\alpha(\theta)$ is a pure super-translation. If we project
onto the pure translational part \cite{Mc1} we can extract the
four momentum $p^\mu=(p_0,0,0,p_3)$ and imposing $p^2_0-p^2_3=0$
we select only the massless particles in the $SO(2)$ description.
Nonetheless we have a completely different behavior from the
Poincar\'e case where the four momentum
$\bar{p}^\mu=(p_0,0,0,p_0)$ is the fixed point of the little group
$E(2)$ thus constraining the massless particle to live on a three
dimensional space generated by $\frac{SL(2,\mathbb{C})}{E(2)}$;
the main reason for this discrepancy is fully encoded in the {\it
pure} super-translational part $\alpha(\theta)$ of the fixed point
which is responsible of the "reduction" of the isotropy group from
$E(2)$ to $SO(2)$. This does not hold for massive $SU(2)$
particles since in this case the fixed point on the orbit for the
BMS group is given by a constant function $p(\theta,\varphi)=m$
and the pure super-translational part is always identically
vanishing.\\
Furthermore, recall that in the Poincar\'e case, the description of massless
particles is rather subtle since a direct application of the
Mackey induction theory associates the physically relevant
massless wave functions to the (unfaithful) representations of the
$E(2)=SO(2)\ltimes T^2$ group. Thus the particles we are
interested in live on the three dimensional orbit
$\frac{SL(2,\mathbb{C})}{E(2)}\sim\Re\times S^2$ whose fixed point
is the four-momentum $p_\mu=(p_0,0,0,p_0)$ but the non compact
nature of $E(2)$ prevents us to write the equations of motion as
an Hilbert space ortho-projection as in the BMS massless case.
Instead we have two options: in the first one we start with a wave
equation living on the orbit of the little group but transforming
under a covariant representation $U^{0,+,(j,j)}$ of
$SL(2,\mathbb{C})$ which implies that the full Hilbert space is
given by the set
\begin{equation}\label{gaugeph}
\mathcal{H}=\left\{\psi:\frac{SL(2,\mathbb{C})}{E(2)}\to\mathbb{C}\;\;|\;\;\
psi\in
L^2(\frac{SL(2,\mathbb{C})}{E(2)})\otimes\mathbb{C}^{2(2j+1)}\right\}.
\end{equation}
This is the space of gauge degrees of freedom and although it is a
symplectic space, we are de facto interested in the space of
physical degrees of freedom which is given by those fields
carrying the $SO(2)$ unitary representations $U^{0,+,j}\oplus
U^{0,+,-j}$ (here $j$ is the value of the helicity for the
massless particle). At the end of the day the physical Hilbert
space is
$$\mathcal{H}_{phys}=\left\{\psi:\frac{SL(2,\mathbb{C})}{E(2)}\to\mathbb{C}\;\
;|\;\;\psi\in
L^2(\frac{SL(2,\mathbb{C})}{E(2)})\otimes\mathbb{C}\right.$$
\begin{equation}
\label{physH}
\left.{\rm and}\;p_{\mu_j}
\psi^{\mu_1...\mu_j}=0\right.\bigg\}/\sim,
\end{equation}
where $\sim$ is the equivalence relation
$$\psi_{\mu_1...\mu_j}\sim\psi_{\mu_1...\mu_j}+p_{\mu_j}\lambda_{\mu_1...\mu
_{j-1}}(p),$$
and where $\lambda(p)$ is a tensor of rank $j-1$.\par
At a level of symplectic phase space, the gauge degrees of freedom can be
modded
out through Marsden-Weinstein symplectic reduction. The advantage of this
approach \cite{bookL} is that we start directly from the symplectic
covariant phase space (\ref{gaugeph}) (endowed with its natural symplectic
form)
and we end up with the space of physical degrees of freedom (\ref{physH})
which
acquires a symplectic structure directly from $\mathcal{H}$.
Thus, to each generator of
$\zeta_a\in\frac{SL(2,\mathbb{C})}{E(2)}$, we can introduce a suitable
hamiltonian for a field in
$\mathcal{H}_{phys}$:
\begin{equation}\label{masslp}
H_{\zeta_a}(\phi)=\frac{-1}{2}\Omega(i\mathcal{L}_{\zeta_a}\phi,\phi)
=\frac{Re}{2}\int\limits_{S^2\times\Re}\frac{d^3p}{p_0}\overline{\mathcal{L}
_{\zeta_a}\phi}(p)\phi(p),
\end{equation}
where $\Omega$ is the symplectic form
on $\mathcal{H}_{phys}$ and $\frac{d^3p}{p_0}$ is the natural
measure on $\frac{SL(2,\mathbb{C})}{E(2)}$.\\
Therefore although both in BMS and Poincar\'e cases the wave
equations transform under the $SO(2)$ group and for this reason
the carrier Hilbert space (i.e. $\mathbb{C}$) is the same, in the
BMS case the particle does not live on the three dimensional
manifold $S^2\times\Re$ but in the five dimensional variety
$S^2\times\Re^3$ embedded in $L^2(S^2)$ and  this is (again!) a
pure super-translational "effect".\par Nonetheless a closer
relation between BMS and Poincar\'e massless particles can be
found if we analyze more in detail the former. In particular, as
we mentioned before, the orbit where the fields live is
constructed applying the elements of the group
$\frac{SL(2,\mathbb{C})}{\Gamma}$ to the fixed point
(\ref{fixedp}) $\bar{p}(\theta,\varphi)\in L^2(S^2)$. Notice that
the function (\ref{fixedp}) has been expanded in spherical
harmonics and that the function $\alpha(\theta)$ consists only on
the sum over all the $Y_{lm}(\theta,\varphi)$ terms with $l>1$.
Let us now consider the map $\pi$ which extracts the Poincar\'e
four momentum from each element in $L^2(S^2)$ i.e.
\begin{equation}\label{proj}
\pi:\mathcal{O}_{BMS}\to\mathcal{O}_P,
\end{equation}
where $\mathcal{O}_P$ is a Poincar\'e orbit. Since the square mass
is a Casimir invariant labelling $\mathcal{O}_{BMS}$, each point
in a BMS orbit is mapped in the same Poincar\'e orbit. In the
specific example of the (double cover of) $SO(2)$ BMS-little
group, the map $\pi$ applied to the fixed point gives the four
momentum $p_\mu=(p_0,0,0,p_3)$ which lies on a massive or massless
Poincar\'e orbit depending on the value of $p_3$. Since we are
interested in massless fields, we choose $p_3=p_0$ which implies
that the BMS orbit for the $\Gamma$ group is fully determined upon the
choice of the energy $p_0$ and of $\alpha(\theta)$.
Moreover, if we remember that for the BMS $SU(2)$ massive little group, the
fixed point was the constant function $p(\theta,\varphi)=m$ thus
with a vanishing super-translation part and that in this scenario
there is a full correspondence between the dynamic of boundary and
bulk massive $SU(2)$ fields, let us consider also in the massless
BMS case the special case of an orbit with vanishing
super-translation i.e. $\alpha(\theta)=0$. Needless to say that,
since there are as many choices as functions $\alpha(\theta)$
invariants under the action of the $SO(2)$ group, we are free to
choose a vanishing super-translational part in the fixed point. In
this case the orbit is generated by the action of
$\frac{SL(2,\mathbb{C})}{\Gamma}$ on
$\bar{p}(\theta,\varphi)=p_0(1+\cos\theta)=p_0[Y_{00}(\varphi)+Y_{10}(\theta
,\varphi)]$
but it is known (see chapter 4.15 in \cite{penroserindler} and
chapter 9.8 in \cite{Penrose2}) that the set of the first four
spherical harmonics transform among themselves under the conformal
motions of the sphere i.e. the group $SL(2,\mathbb{C})$. This
result applies as well for the coset group
$\frac{SL(2,\mathbb{C})}{\Gamma}$ and thus the following relation
holds:
\begin{equation}\label{fix}\frac{SL(2,\mathbb{C})}{\Gamma}\bar{p}(\theta)=
\sum\limits_{l=0}^1\sum\limits_{m=-l}^l
p_{lm}Y_{lm}(\theta,\phi),
\end{equation}
with the mass shell condition
$p_{00}^2=p_{10}^2+p_{11}^2+p_{1-1}^2$. Thus ultimately the orbit
is spanned only by three parameters and the projection map (\ref{proj}) on
the
Poincar\'e orbit is a one to one correspondence which grant us
that (\ref{fix}) identifies the three dimensional variety
$\frac{SL(2,\mathbb{C})}{E(2)}$ embedded in
$\frac{SL(2,\mathbb{C})}{\Gamma}$. \par Thus if we perform the
above particular choice we ultimately end up with a massless
particle living on an orbit $\Re\times S^2$ where the two
translational degrees of freedom in excess respect to the
Poincar\'e scenario have vanished. At level of BMS fields a naive
translation of this reasoning lead us to define for an orbit
labelled by $p_0$ and $\alpha(\theta)=0$ a field
$$\psi:\frac{SL(2,\mathbb{C})}{E(2)}\hookrightarrow\frac{SL(2,\mathbb{C})}
{\Gamma}\to\mathbb{C}\;|\; \psi\in
L^2(\frac{SL(2,\mathbb{C})}{E(2)})\otimes\mathbb{C}.$$ 
In this way we can define a massless field living in the same orbit as the
bulk Poincar\'e field and transforming naturally under a
representation of the $SO(2)$ group without having to mod out any
sort of gauge degrees of freedom. The above discussion from one
side naturally describes in a BMS language the discreteness of the
helicity quantum number for a massless field and from the other
side grants that the Hamiltonian associated to the above function
$\psi$ is by construction identical to (\ref{masslp}).
%%%%%%%%%%%%%%%%%%%%%%%%%%%%%%%%%%%
%%%%%%%%%%%%%%%%%%%%%%%%%%%%%%%%%%%
\section{Concluding remarks}
%%%%%%%%%%%%%%%%%%%%%%%%%%%%%%%%%%%%%
%%%%%%%%%%%%%%%%%%%%%%%%%%%%%%%%%%%%%%%

The philosophy followed in this paper is to assume the existence
of a S-matrix and examine the consequences.\\
We have considered the emergence of the IR sectors in the case of
the gravitational field and their interpretation in term of BMS
representation theory.\\
We have analyzed the extremely complicated mapping bulk-boundary
symmetries, which remains quite not universal. The situation is
definitively much more complicated w.r.t to AdS/CFT, where the
bulk isometry group is exactly the conformal group acting on the
boundary. This in turn
 implies difficulties in building a dictionary bulk-boundary at the level of
fields, due to the plethora(!) of
 boundary fields.\\
We have finally constructed some preliminary building blocks of a
candidate boundary theory invariant w.r.t to the BMS group. We
have seen similarities and differences compared to the Poincar\'e
case and observed that they are connected with purely
super-translation effects in most cases.\\
 An interesting
following step would be to construct the analogue of the OPE for
these fields and see if, at least in a formally, one can have a
definition of a CFT or similar \footnote{As pointed out by the Referee it is the conformal group which is usually
associated with the dual holographic description. An interesting suggestion
along these lines comes from Banks \cite{critique}: the BMS algebra, as
remarked, is the semidirect product of vector fields of the
form $f(\Omega) \partial_u$ with f arbitrary function of the sphere and the
conformal algebra of the sphere. Now, in d=4 the conformal
algebra is nothing else that the infinite dimensional Virasoro algebra. So
Banks conjectures that the correct symmetry algebra of the putative boundary
theory is this large extension of the BMS algebra, or even some subalgebra,
like the one in which $f$ is restricted to the sum of a holomorphic and
anti-holomorphic function on the sphere. We consider this an interesting
proposal but we were not able to make progress along this direction.}. Clearly this theory (or
"structure X" as Witten calls it \cite{barioni}) living on $\Im$
will have some unusual properties due to the degeneracy of the
manifold. It would be also interesting to reconsider our previous
discussions in the super-symmetric case (see \cite{awada} for some
discussion on BMS and SUSY). For the construction of the S-matrix
itself it might be useful to use techniques from twistor\footnote{see
\cite{krasnov} for a related approach.} or
H-spaces \cite{wip}. We leave these topics for future
investigation.

\begin{center}
{\bf Acknowledgments}
\end{center}
We thank P.J.McCarthy for fruitful correspondence on the BMS
representation theory and T. Banks for pointing out ref
\cite{asymptoticquantization}. C.D. wishes to thank M.Carfora and
O.Maj for useful and interesting conversations on dynamical
systems.
 G.A. thanks E.Alvarez for interesting correspondence and explanations of a
related approach \cite{Alvarez} and the members of the High
Energy Theory group at Spinoza Institute (Utrecht) and Racah
Institute (Hebrew University) for interesting questions during
internal seminars where part of this material was presented. The
work of G.A. is supported by a Marie Curie European Fellowship
(Programme "Structuring the European Research Area-Human Resources
and Mobility"). The work of C.D. was supported in part by the
Ministero dell'Universita' e della Ricerca Scientifica under the
PRIN project \emph{The geometry of integrable systems.}
%%%%%%%%%%%%%%%%%%%%%%%%%%%%%%%%%%%%%%%%%%%%%%%%%%%%%%%%%%%%%%%%%%%%%%%%%%%%
%%%%
%%%%%%%%%%%%%%%%%%%%%%%%%%%%%%%%%%%%%%%%%%%%%%%%%%%%%%%%%%
%%%%%%%%%%%%%%%%%%%%%%%%%%%%%%%%%%%%%%%%%%%%%%%%%%%%%%%%%%%%%%%%%%

\end{document}